\documentclass[12pt] {article}
\usepackage{amssymb,amsmath}
\usepackage {natbib} 
\usepackage{color}
\usepackage{hyperref}
\usepackage{graphicx}

\newcommand{\beq}[1]{  \begin{equation} \label{#1} }  	\newcommand{\eeq}{     \end{equation}}  

\renewcommand{\appendix}{ \setcounter{section}{0}\renewcommand{\thesection}{\Alph{section}}
  \section*{Appendix} 
}

\newcommand{\snotz}{s}
\newcommand{\SnotL}{S}
\newcommand{\unota}{\alpha}
\newcommand{\eps}{\epsilon}
\newcommand{\rf}[1]{(\ref{#1})} 
\newcommand{\dd}{{\rm d}} 
\renewcommand{\O}{{\cal O}}
\def\bd#1{\mbox{\boldmath$\displaystyle\mathbf{#1}$} }
\def\Re{\operatorname{Re}} 
\def\Im{\operatorname{Im}} 
\def\sgn{\operatorname{sgn}} 

\def\singlespacing{\baselineskip=13pt}	

\begin{document}   
\date{}                                                                               
\pagestyle{myheadings}\markright{{\sc   Perturbation of a moving crack edge }  }
\singlespacing

\title{ \vskip -0.5in
\textcolor{blue} {A multiple scales approach to crack front waves}}
\author{Andrew N.  Norris\footnote{norris@rutgers.edu}\\
 \it Department of Mechanical \& Aerospace Engineering, \\    
  \it Rutgers University,  Piscataway,  NJ 08854, USA
 \and   
    I. David Abrahams\footnote{i.d.abrahams@manchester.ac.uk}\\  
{\it University of Manchester, Oxford Road, Manchester M13 9PL, UK}   }

\maketitle

\begin{abstract}

Perturbation of a propagating crack with a straight edge is solved using the 
method of matched asymptotic expansions  (MAE).  This provides a simplified analysis in which  the inner and outer solutions are 
governed by distinct mechanics.  The inner solution contains the explicit perturbation 
and is governed by a quasi-static equation. The outer solution determines the radiation 
of energy away from the tip, and requires solving dynamic equations in the unperturbed 
configuration.  The  outer and inner expansions  are matched via the small parameter 
$\eps = L/l$ defined by the disparate length scales: the crack perturbation length 
$L$ and the outer length scale $l$ associated with the loading. The method  is 
illustrated for a scalar crack model and   then applied to the elastodynamic mode I problem.  
The crack front wave dispersion relation is found by requiring that the energy 
release rate is unaltered under perturbation and dispersive properties of the crack front wave speed are described for the first time. 
The example problems considered demonstrate the potential of MAE for moving boundary 
value problems with multiple scales.  

 \end{abstract}

\section{Introduction}

Dynamic perturbation of a steadily travelling crack in a linear elastic medium  is of fundamental interest in fracture  mechanics.  The possible existence of edge supported modulations in an  otherwise straight edge raises questions about the stability of a steadily moving crack front.  These waves,  called  crack front waves, have been the subject of intense scrutiny since they were first observed  by
Rice et al. \cite{Rice94} and Perrin and Rice \cite{Perrin94} through numerical simulations of interactions of dynamic crack fronts with inhomogeneities.  A theoretical framework for crack front waves was provided by Ramanathan and Fisher \cite{Ramanathan97} 
using dynamic weight functions derived earlier by Willis and Movchan \cite{Willis95}.  Ramanathan and Fisher \cite{Ramanathan97} showed that a mode I (opening) disturbance  can propagate  along the crack front with a speed that is a function of the crack velocity and   less than the Rayleigh wave velocity.  Further numerical work confirmed the earlier findings \citep{Morrissey98} and  showed the explicit form of the (non-dispersive) crack front wave speed as a function of the crack speed $v$ 
\citep{Morrissey00,Willis01}.

The theory of Willis, Movchan and Ramanathan has been extended  to other configurations and experimentalists have  sought evidence of  crack front waves.   Woolfries and Willis 
   \cite{Woolfries99} gained insight by examining a scalar crack model.  Woolfries et. al 
\cite{Woolfries02} generalised the  dynamic weight function method to  cracks in viscoelastic materials which was used by Willis and Movchan \cite{Willis01} to show that crack front waves decay in the presence of viscoelasticity.   These theoretical studies considered perturbations of the crack edge in the plane of the moving crack.  Obrezanova et. al
\cite{Obrezanova02} and \cite{Obrezanova02b} examined out-of-plane perturbations of a 2-dimensional crack, and  Movchan et. al 
\cite{Movchan05} considered the dynamic stability of a  crack in a strip. 
Predictions of  crack front waves have also been investigated through experiments.  Crack front waves have been  proposed as the cause of crack surface roughening in brittle materials \cite{Bouchaud02}.  Sharon et. al \cite{Sharon01,Cohen02,Fineberg03}  claim to have evidence of crack front waves in several experiments.  Their conclusions are at odds with those of Bonamy and Ravi-Chandar 
\cite{Bonamy03} who  used ultrasonic shear waves to distort  dynamically growing cracks. They found that the perturbation of the crack front is a 
linear function of the  wave amplitude, but the crack perturbation does not persist after the exciting wave has passed. Many questions on crack front waves remain to be answered. 

The purpose of this paper is to present a new, and, as we will argue, simpler, method for analysing the problem of a perturbed dynamic crack.  The basic idea is that the size of the crack front disturbance is small compared with a macroscopic length scale, which we choose here to be defined by the dynamic loading.  These yield 
inner and   outer length scales and a small parameter $\eps$ which is the ratio of the length scales.  Our approach uses the method of matched asymptotic expansions 
(MAE) to  split the elastodynamic problem into inner and outer sub-problems each of which is simpler than the entire problem and contains the physics appropriate to the region.  As we will demonstrate, the inner problem is quasi-static  and depends explicitly on the assumed form of the crack perturbation.  The outer problem does not consider the fine scale of the perturbation directly, although it determines the radiation of energy from the inner region.   The inner and outer solutions  are related to one another through standard matching arguments.  While the use of asymptotic methods in fracture mechanics is certainly not new, e.g.\ \cite{Willis99},   the method of MAE has not been used  for dealing with complex moving boundary value problems of the type considered here.  MAE is both useful and productive for this class of problem as it naturally splits the problem  mechanically and mathematically.  


The layout of the paper is as follows. 
 We begin in section \ref{sec2} with  a model problem    demonstrating the matched asymptotic approach to studying the perturbation of a travelling crack edge.  The steady solution and the scaling for the perturbation are introduced in section \ref{sec2}, and the details of the MAE analysis are given in section \ref{sec3}. 
  The same approach is then applied in section \ref{sec4} to the physically realistic case of a mode I crack travelling in an elastic material.       The scaling,  matching procedure and  total solution for the elastodynamic case are developed in a manner   analogous to the  scalar problem.    The crack front wave dispersion relation is obtained and new results are presented for the dispersive behaviour of the crack front wave speed,  with  numerical examples given  in section \ref{sec5}  

\section{The scalar model problem of a travelling crack}\label{sec2}

In this section  a model elastic problem of a dynamic  perturbation to a travelling crack front is solved using MAE.    The application is similar to that for the elastic crack but less complex.  We therefore present it in  detail in order to illustrate the general procedure. 

\subsection{Non-dimensional parameters and scaling}

Consider a  function  $\phi(x,y,z,t)$,  motivated by the anti-plane displacement field in mode III elasticity \citep{Rice94}, and  satisfying
a scalar wave equation
\beq{1.1} \nabla^2 \phi - c^{-2} \phi_{tt} = 0,
\eeq
where $\nabla^2$ is the Laplacian operator, the subscript
$t$ denotes partial differentiation in time, and $c$ is the speed of
acoustic waves. A crack occupies
the region
\beq{1.2} x < vt + L f(z,t), \quad y=0, \quad -\infty<z<\infty,
\eeq
that is, it grows at a constant speed, $v$, except for a
perturbation term $L f$, where the length $L$ is the magnitude of the deviation and $f$ is a dimensionless $\O(1)$ quantity.
For purely out-of-plane motions (displacement in the $z$
direction) the crack faces carry zero stress, which we model
here \citep[see][]{Rice94} as
\beq{1.3} \phi_y(x,0,z,t)=0,\quad x<v t+L f,
\eeq
where $\phi_{y} = \partial \phi /\partial y$, etc.
The crack therefore opens at a rate that is approximately constant, and  it is the deviation from the steady state that is of interest.  The opening is assumed to be  
caused by  a steadily translating symmetric loading on the crack faces,
\beq{1.4}
\phi_{y}(x, \pm 0, z, t) =  \frac{P}{l}\, 
p \left(\frac{x-vt}{l}\right) . 
\eeq 
Note that we have assumed that the non-dimensional forcing function $p$ has
argument that scales on a length scale $l$, which is taken to
be {\bf much larger} than the perturbation scale $L$. (For
simplicity we interpret this to also imply that $p(x)$ is zero
in the vicinity of the crack tip.) We define the small parameter $\eps$ as the ratio of the two length scales:
\beq{1.5} \eps = L/l.
\eeq

It is convenient to move to a coordinate system fixed in the
(constant) moving reference frame, and to non-dimensionalise
the initial boundary value problem. Thus, we define
dimensionless independent variables as
\beq{1.6}
\frac{x-vt}{l} \rightarrow x,\quad
\frac{y}{l} \rightarrow y,\quad
\frac{z}{l} \rightarrow z,\quad
\frac{c t}{l} \rightarrow t,
\eeq
together with
\beq{1.7} \frac{v}{c} \rightarrow v,\quad
\frac{\phi}{P}\rightarrow \phi,
\eeq
and the problem reduces to analysing the system
\begin{subequations}\label{18}
\begin{align}
\unota^2 \phi_{xx} + \phi_{yy} + \phi_{z z} - \phi_{tt} + 2 v
\phi_{xt} &= 0, \label{1.8}
\\
\phi_{y}(x, \pm 0, z, t) &=  p(x), \quad x<\eps f(z,t), \label{1.9}
\end{align}
\end{subequations}
where
\beq{1.10}
\unota = \sqrt {1-v^2} \, .  
\eeq
The loading function  $p(x)$ is, by definition, zero close to the crack tip, 
and we assume that the crack growth speed is subsonic so that $\alpha$ remains real.  
The system \rf{18}   still needs to be supplemented by a crack growth criterion, discussed 
below. We  emphasise that here and henceforth all parameters and variables are non-dimensional.

\subsection{Constant running crack}

It will prove useful later to first consider the solution to
\rf{18} for zero perturbation in the crack tip
 position,  $f(z,t)=0$. We start with a
constant running point loading applied symmetrically to the
crack faces at distance unity (i.e.\ at distance $l$ in the
original coordinate system) behind the crack tip:
\beq{1.11}
\phi_{y}(x, \pm 0, z) = \delta(x+1), \quad x<0.
\eeq 
Here $\delta (x) $ is the Dirac delta function. 
Note that, with a symmetric loading, $\phi$ will be zero
ahead of the crack tip, and so this is a standard two-part
boundary value problem 
which may be solved in a variety of ways; a solution of \rf{18}
and \rf{1.11} is therefore
\beq{1.12}
\phi_{y}(x, y, z) = \Re \left\{ \frac{-1}{\pi (x+ i \unota y + 1)
\sqrt{x+ i\unota y}} \right\}.
\eeq
It may be checked, using the identity $(x-i0)^{-1} - 
(x+i0)^{-1} = 2\pi i \delta(x)$, that \rf{1.12} indeed satisfies
\rf{1.11}. Expanding about the crack tip, and integrating, the
solution for $\phi$ is found as
\beq{1.13}
\phi (x, y, z, t) = \Im \left\{ \frac{-1}{\pi \unota}\, \sum
\limits_{n=0}^\infty
\frac{(-1)^n}{n+\frac12 } \left(x+ i \unota y\right)^{n+\frac12} \right\}.
\eeq 
Note that in \rf{1.12} and \rf{1.13}, $\Re$ and $\Im$ denote the
real and imaginary parts of the expressions to their right, and the 
curly brackets are henceforth omitted to maintain clarity. 
More generally, we consider a symmetric
loading along the crack faces such that
\beq{1.14}
\phi_{y}(x, \pm 0, z) = p(x), \quad x<0
\eeq 
with, again, the assumption that $p(x)$ is zero
in a region close to the crack tip.
Based on the line load solution,  the near-tip field is
\beq{1.15}
\phi (x, y, z, t) = \Im P_0
 \left( x+ i \unota y\right)^{\frac12} \bigr[ 1 + \frac{m}{3}
(x+ i \unota y) + \ldots \bigr],
\eeq 
where   
\beq{1.16}
P_0=\sqrt{\frac{2}{\pi}}\,\frac{K_0}{\unota},
\quad 
K_0 = \sqrt{\frac{2}{\pi}}\,\int\limits_{-\infty}^0 \dd s  \frac{ p(s)}{(-s)^{1/2}}, 
\quad m = - \frac{1}{K_0} \,\sqrt{\frac{2}{\pi}}\, \int\limits_{-\infty}^0 \dd s \frac{ p(s)}{(-s)^{3/2}} . 
\eeq
The amplitude $P_0$ defines the strength of the square root 
`displacement' behaviour behind the moving tip
and  $K_0$ is the analogous scalar
`stress intensity factor', such that the stress ahead of the tip is 
\beq{1.17}
\phi_{y} (x, 0, z, t) = \frac{K_0}{\sqrt{2 \pi x }} \,  \big( 1 +m x +
\ldots \big)\, . 
\eeq
 Note that $m = -1$ for the line load limit
\rf{1.11}, and the subscript $0$ in $P_0$
and $K_0$  denotes the displacement and stress coefficients
for the unperturbed crack tip at $x=0$. Finally, the scalar
energy release rate is
\beq{1.18}
G_0 = P_0 K_0
\eeq
where, in general,
\beq{1.19}
P_0 = \lim \limits_{x \uparrow 0} \left[(-x)^{-1/2} \phi (x, 0,
z, t)\right], 
\quad
K_0 = \lim \limits_{x \downarrow 0} \left[\sqrt{x} \, \phi_{y}
(x, 0, z, t) \right].
\eeq

\section{Asymptotic solution; scalar model}\label{sec3}

In the previous section the length scale of the forcing was taken to 
 be much longer than that of the disturbance
to the crack tip. This disparity introduces the small parameter $\eps$
into the problem (in \rf{1.9}) which can  be usefully
employed to solve the problem using the method of matched asymptotic
expansions \citep{Hinch91}. The non-dimensionalisation carried out
previously has cast the initial boundary value problem
into the `outer-field' form so we shall have to derive an
`inner' coordinate system too. In the two coordinate systems,
asymptotic expansions will be deduced, containing sequences
of unknown coefficients, and these will be determined by
matching together the respective terms.

\subsection{The outer expansion}

We begin with the following {\it ansatz} for the outer expansion
\beq{2.7}
\phi (x,y, z,t)= \phi^{(0)} + \eps \phi^{(1)} + \eps^2 \phi^{(2)}
+ \ldots ,
\eeq
where the superscripts in brackets here and henceforth refer
to the expansion function at the order in $\eps$ indicated.
Each term in the expansion must satisfy eq.\ \rf{1.8} as well as
the crack face conditions
\beq{2.71}
\phi^{(n)}_y (x,\pm 0,z,t)=\delta_{n0}\, p(x),\quad x<0, \quad n\ge 0,  
\eeq
where $\delta_{n0}=1$ if $n=0$ and $\delta_{n0}=0$ otherwise.  
The leading order term is the solution for the unperturbed crack, 
\beq{2.8}
\phi^{(0)} =  \Im P_0 \big[ \snotz^{1/2} +\frac{m}{3}
\snotz^{3/2} + \ldots \big]\, ,
\eeq
where the complex variable $\snotz$ is
\beq{2.8a}
\snotz=x+i\unota y. 
\eeq
The complex form of the solution automatically
satisfies a homogeneous crack boundary condition close
to the tip.
Note that,  in terms of the outer coordinates $x$ and $y$  the crack
perturbation is very small, hence the reason for
locating the centre of the coordinate system at the origin.
We will return to the next term in the outer expansion, 
$\phi^{(1)}$, once we have considered the leading terms in the
inner expansion. 
          
\subsection{The inner expansion}
We introduce  dimensionless inner variables $X$ and $Y$,
\beq{3.1}
x=\eps X, \quad y=\eps Y\, , 
\eeq
and let $\Phi (X,Y, z,t) = \phi (x,y, z,t) $.   The wave equation 
expressed in the inner variables is   
\beq{3.2}
\unota^2 \Phi_{XX} + \Phi_{YY} + 2 \eps v \Phi_{Xt}+ \eps^2 \big(
\Phi_{z z} - \Phi_{tt} 
\big)  = 0, 
\eeq
and the edge is now located at $X = f(z, t )$, $Y=0$. 
The inner {\it ansatz} is motivated by the scaling and by the form 
of the leading order outer solution: 
\beq{3.3}
\Phi = \eps^{1/2} \, \Phi^{(1/2)} + \eps^{3/2} \, \Phi^{(3/2)} + 
\ldots . 
\eeq

The leading order term is easily shown to take the form
\beq{3.4}
 \Phi^{(1/2)}  = \Im A^{(1/2)}\SnotL^{1/2},
\eeq
where the inner complex variable $S$ is  
\beq{3.5} \SnotL = X - f + i\unota Y\, ,
\eeq
is centred on the perturbed crack tip in order to capture the correct singular
behaviour at the shifted edge. 
The potential $\Phi^{(1/2)}$ satisfies the governing
equation \rf{3.2} with $\eps$ set to zero (i.e.\ a scaled
form of Laplace's equation) and its derivative is zero on
the crack faces as required.  The unknown $A^{(1/2)}$ is found by 
matching the inner and outer expansions.  More specifically, we 
rewrite the inner expansion in terms of the outer variables.  
Thus, 
taking $\Phi$ up to $\O (\eps^{1/2})$, expanding it in terms of the outer
variables up to $\O (\eps^0)$, and using $\SnotL = \eps^{-1} \snotz -f$,
we obtain
\beq{3.6}
\Phi^{ \{ \frac12 , 0\} } = \Im A^{(1/2)} \snotz^{1/2} \, .
\eeq
Similarly, the expansion of $\phi$ up to $\O (\eps^{0})$ when
expressed in terms of the inner variables up to 
$\O (\eps^{\frac12})$ is
\beq{3.7}
\phi^{\{ 0 ,\frac12\} } = \eps^{1/2} \, \Im P_0 (X+i\unota
Y)^{1/2} \, . 
\eeq
Comparing \rf{3.6} and \rf{3.7}, we see that they are equivalent if $A^{(1/2)} = P_0$.

The next term in the inner expansion satisfies, according to 
\rf{3.2} and \rf{3.3}, 
\beq{3.9}
\unota^2 \Phi_{XX}^{(3/2)} + \Phi_{YY}^{(3/2)} =- 2v
\Phi_{Xt}^{(1/2)}\, .  
\eeq
The solution is a sum of the  particular 
integral and a general solution.  The 
former is readily found using 
$\Phi_{Xt}^{(1/2)}  = \tfrac14 P_0f_t \Im \SnotL^{-3/2}$, 
combined with  the identity 
$(\unota^2 \partial^2_{X}+\partial^2_{Y} ) \bar{\SnotL}
g(\SnotL) = 4\unota^2 g'(\SnotL)$, where $\bar{\SnotL}$ is the
conjugate of $\SnotL$.  Adding the appropriate general solutions, we
determine
\beq{3.11}
\Phi^{(3/2)} = \Im P_0\, \big[ 
A^{(3/2)}\SnotL^{3/2} + B^{(3/2)}\SnotL^{1/2} + \frac{
v}{4\unota^2 } f_t \, \bar{\SnotL}\SnotL^{-1/2} \big]
\eeq
where $A^{(3/2)}$, $B^{(3/2)}$ are as yet unknown.
The expansion of $\Phi$ up to $\O (\eps^{3/2})$ in outer variables
up to $\O (\eps^0)$ is therefore,
\beq{3.12}
\Phi^{ \{ \frac32 , 0\} } = \Im P_0\, \big[  \snotz^{1/2} +
A^{(3/2)}\snotz^{3/2} \big]\, .
\eeq
Similarly, from \rf{2.7} and \rf{2.8}, 
\beq{3.13}
\phi^{ \{ 0, \frac32 \} } =   \Im P_0\, \big[ \snotz^{1/2}
+\frac{m}{3}\snotz^{3/2} \big] .
\eeq
Comparison of the latter two expansions, which should be 
identical 
by the matching rule, implies 
$A^{(3/2)} = m /{3}$.

To summarise, the inner solution has been determined to $\O (\eps^{3/2})$ as follows,
\beq{3.15}
\Phi = \Im P_0 \, \big[ \eps^{1/2} \SnotL^{1/2} +
\eps^{3/2}\big( \frac{m}{3} \SnotL^{3/2} +  B^{(3/2)}\SnotL^{1/2}
+ \frac{ v}{4\unota^2 } f_t \,
\bar{\SnotL}\SnotL^{-1/2}\big) \big] + \O(\eps^{5/2})\, .
\eeq
The single real-valued coefficient  $B^{(3/2)}$ remains unknown, and  will 
be found in the next section.  Finally, we note for future 
reference, 
\beq{3.18}
\Phi^{ \{ \frac32 , 1\} } = \Im P_0 \bigg[\big( \snotz^{1/2}
+\frac{m}{3} \snotz^{3/2} \big)
+\eps \, 
 \big[  - \tfrac{1}{2} f\, \snotz^{-1/2}  + \big(
B^{(3/2)}
-\frac{m}{2} f\big) \snotz^{1/2}
+  \frac{ v f_t}{4\unota^2 } \, \bar{\snotz}\snotz^{-1/2} \big]\bigg]  .
\eeq

\subsection{Wiener-Hopf analysis}
In order to proceed to the next order in the outer solution, which 
is required if we are to match with the inner expansion from 
\rf{3.15}  
it is necessary to consider the perturbation in the 
wavenumber frequency domain.  The most general form of $f(z, t)$
can be constructed from the solution
\beq{4.1} 
f(z, t)  =  \Im f_0 \,e^{i(k z- \omega t)}
=   \Im f_0 \,e^{i\omega (\kappa z- t)},
\eeq
where $\kappa=k/\omega$ (assumed to be less than unity) 
is the edge wave slowness and $f_0$ is a constant.
We seek possible solutions of the outer system of equations with 
no forcing but which  display the singularity represented by the 
$\snotz^{-1/2}$ term in \rf{3.18}.  We therefore assume
\beq{4.2}
\phi^{(1)} (x,y, z,t)= \Im q (x,y)
e^{i\omega (\kappa z- t)}\, ,
\eeq
where $q (x,y)$ satisfies, according to \rf{18},
\beq{4.3}
\unota^2 q_{xx} + q_{yy} + (\omega^2
-k^2) q
-2i\omega v q_{x} = 0, \quad -\infty  <x<\infty, \,
y>0 , 
\eeq
and
\begin{subequations}
\begin{align}\label{4.34}
q_{y}(x,0) &= 0, \quad x<0, \\
 q (x,0) &= 0, \quad x>0, \\
\lim \limits_{x^2+y^2\rightarrow 0} (x^2+y^2)^{1/4} 
q (x,y) &< \infty .
\end{align}
\end{subequations}

One type of general solution is found by a standard analysis 
involving the Wiener-Hopf method.  Thus, let 
\beq{4.5}
q (x,y) =\frac{1}{2\pi} \int\limits_{-
\infty}^\infty \dd \xi\, \hat{q} (\xi)
e^{-i\omega\xi x - \omega \gamma  y }\, , 
\eeq
where $\gamma$ 
follows from \rf{4.3}, 
\beq{4.6a}
\gamma (\xi) = \big( \xi^2 + \kappa^2 -(1 -v\xi)^2\big)^{1/2}.
\eeq
The integration contour runs along a strip in the complex
$\xi$ plane, $\cal D$ say, which contains the real line except
that it is indented above the point $-\lambda_+$ and below
the point $\lambda_-$, where these correspond to the branch points
of $\gamma$. The Riemann surface of $\gamma$ is selected so that
$Re\, \gamma \ge 0$. 
We further define
\beq{4.6}
\gamma = \gamma^+ \gamma^-\, , 
\eeq
with
\beq{4.7}
 \gamma^{\pm}(\xi) = \unota\, \big( \xi \pm \lambda_{\pm}\big)^{1/2}, \qquad
\lambda_{\pm} \equiv  \frac{1 }{\unota^2}  \big(
1-\kappa^2 \unota^2\big)^{1/2}
\pm \frac{ v}{\unota^2} \, .
\eeq
By definition, the function $\gamma^+$ $(\gamma^-)$ is analytic
in the upper (lower) half plane containing the common strip of
regularity $\cal D$.
Generally, the superscript notation $\pm$ indicates
henceforth functions analytic in these overlapping half-planes\footnote{However, the reader is reminded that the subscript on the
constants $\lambda_\pm$ refers only to the choice of sign in \rf{4.7}!}. 
The boundary conditions on $x<0$ and $x>0$ imply, respectively, 
\begin{subequations}
\begin{align}\label{4.8}
\gamma \hat{q} (\xi)  &= T^+ (\xi) ,\\
\hat{q} (\xi)  &= W^- (\xi) .
\end{align}
\end{subequations}
Thus  $\gamma^- W^- (\xi)$ and $ T^+ /\gamma^+$
are equal to one another 
and hence by analytic continuation and Liouville's 
theorem must be also equal to a constant, say 
$q_0$. 
Therefore, $\hat{q}  (\xi) = q_0/\gamma^- (\xi)$.

The near tip behaviour of $q (x,y)$ follows from the
behaviour of the transform at large $\xi$, thus, as $r\rightarrow 0$, 
\beq{4.10}
q (x,y) =\frac{q_0}{2\pi\alpha} \int\limits_{-
\infty}^\infty \frac{\dd \xi\, e^{-\omega(i\xi x + |\xi| \unota y)}}{ (\xi -
i0)^{1/2} } \, 
\bigg[ 1 + \frac{\lambda_- }{2\xi }
-\frac12 \unota y (\lambda_+-\lambda_-) \sgn  \xi
+\unota y (\lambda_++\lambda_-)^2\, \frac{\sgn  \xi}{8\xi}
+ \ldots \bigg]. 
\eeq    %
This can be evaluated using the identities 
\beq{4.11}
\frac{1}{2\pi} \int\limits_{-\infty}^\infty \frac{\dd \xi\, e^{-\omega(i\xi x + |\xi| \unota y)} }{(\xi -i0)^{1/2} \, \xi^{n}}\,  \bigr[1,\,
\sgn  \xi\bigr]
 = 
-  \frac{2^{2n}n!}{(2n)!} \frac{e^{i\pi/4} }{\sqrt{\pi}} (-
i)^n\,|\omega\snotz|^{n-\frac12}\, 
\bigr[ \sin (n-\tfrac{1}{2} )\theta\, , i \cos (n-\tfrac{1}{2} 
)\theta \bigr],
\eeq
for $n\ge 0$, 
where 
$\snotz=x+i\unota y \equiv |\snotz| e^{i\theta}$. Hence,
\begin{align}\label{4.12}
q (x,y) =& \frac{q_0}{\alpha} \frac{e^{i\pi/4} }{\sqrt{\pi\omega}}
\bigg( |\snotz|^{-1/2} \sin \frac{\theta }{2}
 + i \omega\lambda_- |\snotz|^{1/2} \sin \frac{\theta }{2}
\nonumber \\
& + \big[ \frac{i}{2} \omega(\lambda_+-\lambda_-)|\snotz|^{1/2} - \frac{\omega^2}4 (\lambda_++\lambda_-)^2
|\snotz|^{3/2} \big] \sin\theta \cos \frac{\theta }{2}
+ \ldots \bigg)\, . \quad 
\end{align}
Thus, from  \rf{4.2}, 
the second order term in the outer expansion is 
\beq{4.13}
\phi^{(1)} = \Im   \frac{q_0e^{i\frac{\pi}4} }{\alpha\sqrt{\pi\omega}}
\bigg( |\snotz|^{-\frac12} \sin \frac{\theta }{2}
 + \frac{i}{4}\omega |\snotz|^{\frac12} \big[(\lambda_+ +3\lambda_-) \sin \frac{\theta }{2}
+   (\lambda_+-\lambda_-)\sin \frac{3\theta }{2}\big] 
 \bigg)e^{i(k z- \omega t)}\, + \mbox{O}( |\snotz|^{\frac32}) \, .
\quad 
\eeq

We are now ready to complete the matching with the inner field $\Phi^{ \{ \frac32 , 1\} }$ of 
\rf{3.18}.  This requires that the coefficients of 
$|\snotz|^{-1/2} \sin \frac{\theta }{2}$, $|\snotz|^{1/2} \sin
\frac{\theta }{2}$ and $|\snotz|^{1/2} \sin \frac{3\theta }{2}$,
are identically equal.  The former implies that 
\beq{987}
q_0  = \tfrac12\alpha \sqrt{\pi\omega}e^{-i\pi/4}  P_0 f_0, 
\eeq
while the $|\snotz|^{1/2} \sin \frac{\theta }{2}$ terms match if
\begin{align}\label{4.15}
B^{(3/2)}&= \Im    \bigr[ \frac{m}2 + \frac{i}{8}\omega (\lambda_+ +3\lambda_-)
\bigr] f_0e^{i(k z- \omega t)}\, . 
\nonumber \\
&= \Im    \bigr[ \frac{m}2 + \frac{i}{2\unota^2} \big( \omega^2 - k^2 \unota^2\big)^{1/2} 
\bigr] f_0e^{i(k z- \omega t)} + \frac{ v}{4\unota^2 } f_t, 
\end{align}
where $\rf{4.7}_2$ has been used to express it in a form that will be useful later. 
Finally, we note that the  $|\snotz|^{1/2} \sin \frac{3\theta }{2}$ terms 
automatically match on account of the identity $(\lambda_+ - \lambda_-) = 2 v /\unota^2$.  We are now ready to consider  the complete MAE solution.

\subsection{Solution and discussion}
The strengths of the perturbed singularities at the perturbed crack edge are defined by 
\begin{subequations}
\begin{align}
P &\equiv \eps^{-1/2}\lim \limits_{ X \uparrow 0} |X |^{-1/2} \Phi (X,0, 
z, t)
=  P_0 + \eps P_0\big( B^{(3/2)} + \frac{ v}{4\unota^2 } f_t \big), \label{5.2}
\\
K &\equiv \eps^{-1/2}\lim \limits_{ X \downarrow 0} X^{1/2} \Phi_Y (X,0, 
z, t)
=  K_0 + \eps K_0 \big(  B^{(3/2)} - \frac{3 v}{4\unota^2 } f_t \big), \label{5.3}
\end{align}
\end{subequations}
where the values follow from the  leading order terms in the inner solution in \rf{3.15} using \rf{4.15}.
Consider the variations from the  values for the steadily propagating crack: $\Delta P = P-P_0$, $\Delta K = K-K_0$ and $\Delta G = G-
G_0$, where  $G=PK$ is the perturbed energy release rate, then the relative changes are 
\beq{5.5}
\frac{\Delta P }{P_0}= \eps g*f + \eps \frac{vf_t}{2\alpha^2}, \quad \quad 
\frac{\Delta K }{K_0}= \eps g*f - \eps \frac{vf_t}{2\alpha^2}, \quad \quad  \frac{\Delta G 
}{G_0} = \eps g*f, 
\eeq
where $*$ denotes convolution, and the transform of  $g(t)$ is
\beq{5.6}
\hat{g}(\omega, k) =  m + i\unota^{-2} \big( \omega^2 - k^2 \unota^2\big)^{1/2}  \, .
\eeq
The expressions for $\Delta K $ and $\Delta G$ are identical to 
analogous ones derived by Woolfries and  Willis (1999) (eqs.\ 
(2.5), (2.7) and (2.18) of their paper).  They also agree with 
prior work by Rice et al.\ (1994) for the special case of $m = 0$.

Crack front waves are possible if the phase 
speed $c_p = \omega /k$ and the crack speed $v$ together lie inside the sonic cone, i.e.\ they satisfy 
\beq{555}
c_t^2\equiv v^2 + c_p^2 < 1. 
\eeq  
In that case the transform $\hat{g}$ becomes
\beq{5.61}
\hat{g}(\omega, k) =  m- |k|  \frac{\sqrt{1-c_t^2}}{1-v^2} \, .  
\eeq
Zeros are possible only if  $m$ is positive and satisfies 
\beq{5.62}
0<m <  |k|/\sqrt{1-v^2}, 
\eeq
 in which case $ \hat{g}(\omega, k)$ has a unique zero at 
$\omega = \unota \sqrt{k^2 - \unota^2 m^2}$.  
This  solution has been discussed before, but we note briefly some 
properties in the wavenumber domain. 
The phase speed as a function of $k$  is monotonically 
increasing from zero at the cutoff or lowest wavenumber possible, 
$k =  m\sqrt{1-v^2} $, to its asymptote $c_p=\sqrt{1-v^2}$ as $k \rightarrow \infty $.  The crack edge waves are therefore dispersive, and one can, formally at least, define 
a  group  speed (velocity),  $c_g \equiv d \omega /dk$.  This is related to the phase speed by 
\beq{5.10} 
c_p c_g = 1-v^2, 
\eeq
and hence, for finite frequency/wavenumber 
\beq{5.101} 
v^2 +c_g^2 >1.  
\eeq
This type of dispersion is anomalous in that the group speed is normally associated with the energy propagation speed.   However, there is no strict connection, and in fact, the notion of energy propagation for edge waves has not been defined.  As a result, we may define and calculate $c_g$ although its physical significance is not entirely clear.  

It is also of interest to examine the final solution for the near field, 
as it   provides the complete behaviour in the 
neighbourhood of the moving crack front.   Equation \rf{3.15} implies that the inner solution  
 can be expressed  
\beq{5.8}
\Phi = P_0\bigg( 
\big[ 1+ \eps \big(\frac12 g*f + \frac{ vf_t }{4\alpha^2} \big) \big]\rho^{1/2}\sin \frac{\Theta 
}{2} 
+\big[ \frac{m}{3} \rho^{3/2} - \eps \frac{ vf_t}{4\alpha^2}   \rho^{1/2}
\big] 
\sin \frac{3\Theta}{2} \bigg) + \O (\eps^{3/2}), 
\eeq
where $\rho, \Theta$ are polar coordinates relative to the 
perturbed moving tip: $x - \eps f + i\unota y = \rho e^{i\Theta}$.
We note the appearance of the $\rho^{1/2}\sin \frac{3\Theta}{2} $ term, which 
affects the stress singularity ahead of the crack, but has a 
different angular dependence from the standard near tip field 
$\rho^{1/2}\sin \frac{\Theta }{2} $ for a steadily moving crack. 

\subsection{Properties of  the MAE scheme and simplification}
Before considering the elastic crack in detail, we note some general features of the matched asymptotic analysis for the scalar problem.  The sequence of 
terms derived was 
$\phi^{(0)}$ of \rf{2.7} $\rightarrow \Phi^{(1/2)}$ of \rf{3.3} $\rightarrow \Phi^{(3/2)}$ of \rf{3.15}
$\rightarrow \phi^{(1)}$ of \rf{4.13}. The  latter was derived  as an eigensolution using the Wiener-Hopf method, and was necessary in order  to complete the matching of the  single remaining coefficient in $\Phi^{(3/2)}$.   Note that the particular solution of the inner term for the forcing in eq.\ \rf{3.9} was not required for  matching.  In fact, the terms in $\Delta P$ and $\Delta K$  that involve $f_t$ cancel in the final expression for  $\Delta G$, i.e.\ $\hat{g}$ of
\rf{5.6}.  

Thus, we calculated the perturbed values of $P$ and $K$ separately, although the quantity of interest, $G$, which is their product, turns out to be simpler.  This suggests a more direct procedure, which  follows by noting that $G$ has alternative expressions: 
\beq{66}
G =PK = \sqrt{\frac{\pi}{2}} \, (1-V^2)^{-1/2}\, K^2  = \sqrt{\frac{2}{\pi}}\,(1-V^2)^{1/2}\, P^2 , 
\eeq
where 
$V = v + \eps f_t$ is the velocity of the crack edge.  
The identities \rf{66} are  a consequence of the general form of $\rf{1.16}_1$ for speed $V$, 
 \beq{67}
K=\sqrt\frac{\pi}{2}\,(1-V^2)^{1/2}\, P . 
\eeq
As a consequence the perturbed energy release rate can be expressed in different ways,   
\beq{68}
\frac{\Delta G}{G_0}  = \frac{\Delta P}{P_0} + \frac{\Delta K}{K_0} 
= 2\frac{\Delta K}{K_0} +\frac{v}{\unota^2}f_t = 2\frac{\Delta P}{P_0}  -\frac{v}{\unota^2}f_t. 
\eeq
Only one or other of $P$ or $K$ needs to be considered in order to calculate the energy release rate.  In practice, it is simpler to compute $P$ as it is the field variable calculated in  the MAE procedure.    Restricting attention to the crack face $X<0,\, Y=+0$, and using eqs.\ \rf{3.15} and \rf{3.18}, we have 
\beq{3.182}
\Phi(-|X|,+0,z,t)= P_0 \big[\eps^{1/2} (1+\eps p_1) |X|^{1/2} -
\eps^{3/2} \frac{m}{3} |X|^{3/2} \big]  + \O(\eps^{5/2}),
\eeq
and
\beq{3.183}
\Phi^{ \{ \frac32 , 1\} } (-|X|,+0,z,t)=   P_0 \bigg[\big( |x|^{1/2}
-\frac{m}{3} |x|^{3/2} \big)
+\eps \,  \big[   \tfrac{1}{2} f\, |x|^{-1/2}  + \big(
p_1 -\frac{m}{2} f\big) |x|^{1/2}  \big]\bigg]  ,
\eeq
where 
\beq{747}
p_1 = B^{(3/2)} + \frac{vf_t}{4\alpha^2} = \eps^{-1} \frac{\Delta P}{P_0}. 
\eeq
At the same time, the outer expansion evaluated on the crack face is, from \rf{3.13}, \rf{4.2} and \rf{4.10},  
\beq{384}
\phi^{ \{1, \frac32\} } =  P_0 \big[ |x|^{1/2} -\frac{m}{3} |x|^{3/2} 
+\eps \, \Im e^{i(kz-\omega t)} 
\frac{q_0}{2\pi\alpha} \int\limits_{-
\infty}^\infty \frac{\dd \xi\, e^{ i\omega \xi |x| }}{ (\xi -
i0)^{1/2} }  
\, \big( 1 + \frac{\lambda_- }{2\xi } +\O(\xi^{-2}) \big)
\big].
\eeq
The integral can be evaluated using \rf{4.11}. 
Matching the $\O(\eps)$ terms in \rf{3.183} and \rf{384},  the $|x|^{-1/2}$ singularity gives the identity \rf{987} for $q_0$, while the next term yields
\beq{600}
p_1 - \frac{m}2 f = 
\Im    \, i\omega \frac{\lambda_-}{2}  f_0e^{i(k z- \omega t)}. 
\eeq
The perturbed energy release rate then follows from eqs.\ \rf{68}, \rf{747} and \rf{600}. 

In summary, we have presented two methods to complete the matching - the first and more general applies the matching to  the field $\phi (x,y,z,t)$ for all $x$ and $y$ near the origin, and the second  only matches the field on the crack faces, $X<0,\, Y=0$.  The two methods are equivalent because the final stage of the matching requires only a single real-valued coefficient,  and therefore matching of the field along a line amounts to matching over an area (in $x$ and $y$).  However, $B^{(3/2)}$ is not the most relevant quantity for the purpose of calculating the perturbation in energy release rate, although this coefficient can be found using the second approach, from eq.\ \rf{747} once the coefficient $p_1$ is known. 
These same general features are repeated in the elastic case. In particular the calculation of the perturbed energy release rate will be achieved directly once the matching is completed. 

\section{The elastic crack} \label{sec4}

We now turn to the more realistic problem of  
 in-plane perturbations of a steadily propagating mode I crack in an isotropic elastic solid.  
It is important to distinguish between two distinct types of `crack edge waves'.  The first is the 
analogue of the Rayleigh
surface wave, that is, an infinitesimal wave of particle 
displacement confined to the edge and decaying away from the edge.  
The second is a wave-like modification of the edge itself.   
Achenbach and Gautesen 
\cite{Achenbach77} 
demonstrated the nonexistence of the first type of  edge wave on a 
stationary crack edge.  It can be shown using their arguments  combined with the  analysis in the Appendix, that this result extends to 
 to a steady  propagating crack.  That is, there 
there are no localised solutions of this form for any crack 
velocity that is subsonic relative to the Rayleigh wave velocity.  

\subsection{Scaling and asymptotic expansions}

The unperturbed crack  lies in the plane $y=0$, with  infinite 
edge moving steadily  at speed $v$,   located at $x=0$, $-\infty 
< z < \infty$,  in the convected coordinate system.  
We adopt the same scaled coordinates as in the scalar problem, with the distinction that 
the normalisation in time is with respect to the Rayleigh wave 
speed $c_R$, so that the non-dimensional variables are 
\beq{7.1}
\left( \frac{x-vt}{l}, \frac{y}{l}, \frac{z}{l}, \frac{c_R}{l}t \right) \rightarrow (x, y, z, t), 
\qquad \frac{v}{c_R} \rightarrow v. 
\eeq
The inner and outer expansions for the  displacement field ${\bd u} = ( u_x, u_y, u_z )^T$ are 
\begin{subequations}\label{202}
\begin{align}\label{55}
{\bd u}(x,y,z,t) &= {\bd u}^{(0)} + \eps  {\bd u}^{(1)} + \eps^2  {\bd u}^{(2)} + \ldots , 
\\
{\bd U}(X,Y,z,t) &= \eps^{1/2}{\bd U}^{(1/2)} + \eps^{3/2}{\bd U}^{(3/2)} +  \ldots , \label{55b}
\end{align}
\end{subequations}
where the inner variables $X$ and $Y$ are as the same as before, eq.\ \rf{3.1}.  The expansions in \rf{202} are the analogs of those in eqs.\ \rf{2.7} and \rf{3.3} for the scalar problem.  
The boundary conditions on the crack faces are that the traction vector $\bd{\sigma} = (\sigma_{xy}, \sigma_{yy}, 
\sigma_{zy})^T$ vanishes, except for the loading, defined below.

We introduce the  non-dimensional speeds associated with the two bulk wave speeds in an isotropic elastic medium
\beq{131}
 v_I =  {c_I}/{c_R}, \quad I=L,\, T,
 \eeq
where $c_L=\sqrt{(\lambda+2\mu)/\rho_0}$ and $c_T=\sqrt{\mu /\rho_0}$  are the longitudinal  and transverse  wave speeds,  $ \lambda $ and $\mu$ are the Lam\'e moduli and $\rho_0$ is the mass density. 
The displacement field is represented by three potentials:
\beq{6.1}
{\bf u} = \nabla \phi_L + \nabla \wedge \phi_T {\bf e}_3 +  \nabla \wedge \nabla \wedge \phi_{TH} {\bf e}_3, 
\eeq
where the $L$ and $T$ are associated with the longitudinal  and 
transverse  waves, respectively, and $TH$ is transverse horizontal.  
The second transverse potential, $\phi_{TH}$, is zero in the absence of the perturbation.  The effect of changing to the moving coordinate $x $ is that each  $\phi_I$ satisfies a modified wave equation
\beq{1.07}
\unota_I^2 \phi_{I, xx} +\phi_{I, yy} +\phi_{I, zz} - \frac{1}{v_I^2} \, \phi_{I, tt} 
+ 2\frac{v}{v_I^2} \, \phi_{I, xt} = 0 , \quad I=L,T,TH
\eeq
with  
\beq{6.15}
 \unota_I    \equiv \unota_I  (v) = \sqrt{1-  {v^2}/{v_I^2}}, \quad I=L,T,
\eeq
and $\unota_{TH} = \unota_T$,  $v_{TH} = v_T$.   The asymptotic analysis will be performed in terms of the potentials, represented as a 3-vector for both the   outer and inner regions: 
\beq{56}
\bd{ \phi} \equiv 
\begin{pmatrix}
\phi_L \\  
\phi_T\\  
\phi_{TH}
\end{pmatrix} ,
\qquad
\bd{ \Phi} \equiv 
\begin{pmatrix}
\Phi_L \\  
\Phi_T\\  
\Phi_{TH}
\end{pmatrix} , 
\eeq
respectively, and which are expanded as 
\begin{subequations}\label{72}
\begin{align} 
\bd{\phi} (x,y, z,t) &= \bd{\phi}^{(0)} + \eps \bd{\phi}^{(1)} 
+ \eps^2 \bd{\phi}^{(2)} + \ldots , \label{7.2}
\\
\bd{\Phi} (X,Y, z,t) & = \eps^{1/2} \bd{\Phi}^{(1/2)} + \eps^{3/2} 
\bd{\Phi}^{(3/2)} + \ldots . \label{7.4}
\end{align}
\end{subequations}

\subsection{Steadily propagating crack}
We first consider solutions for the steadily propagating crack in an isotropic elastic medium. 
The unperturbed solution has $\phi_{TH}=0$ and the potentials $\phi_L$ and $\phi_T$  may be represented by analytic functions  of two complex variables  
\beq{6.2}
\snotz_I = x+ i\unota_I y, \qquad  \quad I=L,\, T.  
\eeq
We introduce  the real valued potentials,  $\bd{ \psi}_n  $, which we express as functions of the complex variables $\snotz_L$, $ \snotz_T$,
\beq{6.3}
\bd{ \psi}_n (\snotz_L, \snotz_T) \equiv
\frac{1}{\mu \sqrt{2\pi} D (n+\frac32 )  (n+\frac12 )}\, \Re 
\begin{pmatrix}  (1+\unota_T^2) \, \snotz_L^{n+\frac32}
\\  \\
i2 \unota_L\, \snotz_T^{n+\frac32}
\\  \\
0
\end{pmatrix} ,   
\eeq
where 
\beq{6.4}
D (v) = 4\unota_L\unota_T - (1+\unota_T^2)^2, 
\eeq
and $D(1)=0$ based on the normalisation of the wave speeds with respect to the Rayleigh speed. 
The associated vectors of displacement and traction  are 
\begin{subequations}
\begin{align}\label{6.5}
{\bf \Upsilon}_n
&= \frac{1}{ (n+\frac12 )\mu \sqrt{2\pi} D}\, \Re 
\bigg[
(1+\unota_T^2)\, \snotz_L^{n+\frac12}
\begin{pmatrix}
1 \\   i\unota_L \\ 0 
\end{pmatrix} 
+ i2 \unota_L\, \snotz_T^{n+\frac12}
\begin{pmatrix}
i\unota_T \\   -1 \\ 0 
\end{pmatrix}
\bigg] ,  \quad 
\\         
\bd{ \Sigma}_n
&= \frac{-1}{ \sqrt{2\pi} D}\, \Re  
\bigg[  
(1+\unota_T^2)\, \snotz_L^{n-\frac12}
\begin{pmatrix}
-2i\unota_L \\   1+\unota_T^2 \\ 0 
\end{pmatrix}
+ i2 \unota_L\, \snotz_T^{n-\frac12}
\begin{pmatrix}
 1+\unota_T^2\\   2i\unota_T \\ 0  
\end{pmatrix}
\bigg] .
\end{align}
\end{subequations}
We note in particular, that the zero traction conditions on the crack face are satisfied. 

These fundamental elements may be used to describe a steadily propagating mode I crack, with zero applied shear on the crack faces, 
i.e.\ $\sigma_{xy}(x,\pm 0) = \sigma_{zy}(x,\pm 0) =  0$ for $x<0$.  
For instance, the solution for a  pair of travelling line loads 
with  $\sigma_{yy}(x,\pm 0) = -\delta(x+1)$ is 
\beq{6.7}
\sigma_{yy}(x,y) = \frac{1}{  \pi D}\, \Re  
\left[ \frac{4\unota_L\unota_T}{(1+\snotz_T)\snotz_T^{1/2} }
- \frac{ (1+\unota_T^2)^2}{(1+\snotz_L)\snotz_L^{1/2} }\right]\, .
\eeq
The near tip fields, 
in this case, can be expressed  terms of the basis functions explicitly as 
\beq{6.8}
{\bd u}  = \sqrt{\frac{2}{\pi  }} \sum\limits_{n=0}^\infty 
(-1)^n {\bf \Upsilon}_n, \qquad
{\bd \sigma } = \sqrt{\frac{2}{\pi  }} 
\sum\limits_{n=0}^\infty (-1)^n \bd{ \Sigma}_n\, . 
\eeq
More generally, the loading 
\beq{6.9}
\sigma_{yy}(x,\pm 0) = -p(x), \quad x<0, 
\eeq
implies a near tip stress expansion  identical to that for the scalar problem (see \rf{1.17})
\beq{6.10}
 \bd{ \phi}  = K_0 \, \big(
\bd{ \psi}_0 + m \bd{ \psi}_1 + \ldots \big)   
\quad \Leftrightarrow \quad
\sigma_{yy}(x, 0) = \frac{K_0H(x)}{\sqrt{2 \pi x }} \,  \big( 1 +m x +
\ldots \big)\, .
\eeq
Here $K_0 $ is the stress intensity factor, $H$ is the Heaviside (step) function, and 
\beq{6.11} 
K_0 = \sqrt{\frac{2}{\pi}}\,\int\limits_{-\infty}^0 \dd s \, (-s)^{-1/2} \, p(s),\quad 
\quad m = - \frac{1}{K_0} \,\sqrt{\frac{2}{\pi}}\, \int\limits_{-\infty}^0 \dd s \, (-s)^{-3/2} \, p(s)  . 
\eeq
This solution serves as the basis for the perturbed crack.

\subsection{Perturbed crack: inner and outer analysis}

The leading order term in the outer expansion \rf{7.2}   is, from \rf{6.10},  
\beq{7.3}
\bd{ \phi}^{(0)} = K_0 \, \big[ 
\bd{ \psi}_0 (\snotz_L,\snotz_T) +m \bd{ \psi}_1(\snotz_L,\snotz_T) + \mbox{O}(\snotz_L^{3/2},
\snotz_T^{3/2}) \big] \, .
\eeq

Turning to the  the inner expansion we note that the 
  potentials satisfy modified wave equations
\beq{7.36}
\unota_I^2 \Phi_{I,XX} + \Phi_{I,YY} + \eps 2\frac{v}{v_I^2}
\Phi_{I,Xt}+ \eps^2 \bigr( \Phi_{I, z z} - \frac{1}{v_I^2}\Phi_{I, tt}
\bigr)  = 0, \quad I=L,\, T,\, TH. 
\eeq
The leading term in the inner expansion \rf{7.4} follows from \rf{7.3} as  
\beq{7.5}
\bd{\Phi}^{(1/2)} = K_0 \, \bd{ \psi}_0 (\SnotL_L,
\SnotL_T),
\eeq
where the inner variables are 
\beq{7.6} \SnotL_I = X - f + i\unota_I Y\, , \quad I=L, \, T.
\eeq
The potentials defined by \rf{7.5} satisfy the governing equations \rf{7.36}, and the traction conditions on the crack faces.  

The potentials in the next term of the inner expansion satisfy the inhomogeneous wave 
equations
\beq{7.7}
\unota_I^2 \Phi_{I, XX}^{(3/2)} + \Phi_{I, YY}^{(3/2)} =- 2\frac{v}{v_I^2}\,  \Phi_{I, Xt}^{(1/2)}\, , \quad I=L, \, T, \, TH.
\eeq
These may be solved in the same manner as before, as a sum of homogeneous solutions plus the particular solution, 
\beq{7.9}
\bd{ \Phi}^{(3/2)} = K_0 \, \big[ 
A^{(3/2)}\bd{ \psi}_1 (\SnotL_L,\SnotL_T) 
+B^{(3/2)}\bd{ \psi}_0 (\SnotL_L,\SnotL_T)  + vf_t  \bd{ \psi}_*(\SnotL_L,\SnotL_T)\big] , 
\eeq
where the final term is the particular solution 
given by  
\beq{79}
 \bd{ \psi}_*(\SnotL_L,\SnotL_T) = \frac{1}{ \mu \sqrt{2\pi} D}\, \Re 
\begin{pmatrix}
\frac{(1+\unota_T^2)}{ \unota_L^2v_L^2}  \, \bar{\SnotL}_L \SnotL_L^{1/2}  \\    
\frac{ 2i\unota_T}{ \unota_T^2v_T^2} \, \bar{\SnotL}_T \SnotL_T^{1/2}  \\   0  
\end{pmatrix}
 \, .
\eeq
Matching the outer term ${\bd \phi}^{ \{ 0, \frac32 \} }$ with 
${\bd \Phi}^{ \{  \frac32 , 0\} }$   implies that $A^{(3/2)} = m$. 

The complete inner solution is therefore
\beq{799}
\bd{ \Phi} = K_0 \, \big[ \eps^{1/2} \bd{ \psi}_0 (\SnotL_L,\SnotL_T) + 
\eps^{3/2} \big( 
m\bd{ \psi}_1 (\SnotL_L,\SnotL_T) 
+B^{(3/2)}\bd{ \psi}_0 (\SnotL_L,\SnotL_T)  + vf_t  \bd{ \psi}_*(\SnotL_L,\SnotL_T)\big)\big] 
+ \O(\eps^{5/2}), 
\eeq
 from  which we can express the  expansion of the inner in terms of the outer variable as
\begin{align}\label{7.12}
\bd{ \Phi}^{ \{  \frac32 , 1\} } = K_0 \, \bigg( 
& \bd{ \psi}_0 (\snotz_L,\snotz_T) +m \bd{ \psi}_1(\snotz_L,\snotz_T) +  \eps
\bigr[\frac12 f \bd{ \psi}_{-1}(\snotz_L,\snotz_T)
\nonumber \\ 
& + \big( B^{(3/2)} - \frac12 mf\big) \bd{ \psi}_0 (\snotz_L,\snotz_T)  
+vf_t  \bd{ \psi}_*(\snotz_L,\snotz_T) \bigr]\bigg)\, .
\end{align}
These two equations are  the equivalents of eqs.\ \rf{3.15} and \rf{3.18} for the perturbed scalar  crack model.  We note that, as in the scalar problem, the issue is reduced to finding a single real constant, $B^{(3/2)}$. 

\subsection{Matching and final solution}
Based on the experience with the scalar problem, we assume that the $\O(\eps)$ outer solution is time-harmonic and   of the form 
\beq{1.05}
{\bd \phi}^{(1)}  = \Re e^{i\omega (\kappa z -t)}\, 
\frac{\omega}{2\pi}
\int_{-\infty}^\infty \dd \xi  \, e^{-i\omega \xi x  }\, 
\hat{\bd \phi}^{(1)} (\xi, y),
\eeq
where  
\beq{39}
\kappa = 1/c_p
\eeq
 denotes the  slowness  for an assumed edge disturbance with 
phase speed   $c_p$.  The crack edge perturbation itself is defined by \rf{4.1}.  The transform $\hat{\bd \phi}^{(1)} (\xi, y)$ is derived in the Appendix, to within a multiplicative factor.  The latter is determined by the near-tip expansion, which is expressed via the large $|\xi|$ behaviour of the transform.  Thus, from the Appendix, we have 
\begin{align}\label{402}
{\bd \phi}^{(1)}  =& \Re e^{i\omega (\kappa z -t)}\, 
\frac{\omega}{2\pi}
\int_{-\infty}^\infty \dd \xi  \, e^{-i \xi x  }\, \frac{b_0}{\xi^{3/2}}
\begin{pmatrix}
-i(1+\unota_T^2)  e^{-|\xi|\alpha_L y} (\sgn \xi) \big( 1 +    \O (\xi^{-1})\big) 
\\ \\ 
-2\unota_L e^{-|\xi|\alpha_T y}  \big( 1 + \O (\xi^{-1})\big) 
\\ \\
\O (\xi^{-2})
\end{pmatrix} 
\nonumber \\
=& \Re e^{i\omega (\kappa z -t)}\, b_0 \frac{e^{-i\pi/4}}{\sqrt{2}} \, \mu D \, 
\big[\bd{ \psi}_{-1}(\snotz_L,\snotz_T)+ \O(\snotz_L^{3/2},\snotz_T^{3/2})
\big]  ,
\end{align}
where the identities \rf{4.11} have been used. 
  The coefficient $b_0$ is found by matching the coefficient of 
$\bd{ \psi}_{-1}$  with the $\O (\eps)$ 
  term    in \rf{7.12}, to give 
\beq{404}
  b_0 = \frac{e^{-i\pi/4}}{\sqrt{2}} \frac{K_0f_0}{\mu D}. 
\eeq
Hence, the near tip expansion is
\beq{932}
{\bd \phi}^{(1)}  = \frac{f}{2}K_0 \bd{ \psi}_{-1}(\snotz_L,\snotz_T)
+  \O(|x|^{3/2},|y|^{3/2}). 
\eeq
The next order term can be evaluated using the explicit expressions in the Appendix, and in principle the near tip expansion can be continued.  However, the leading order term suffices for our present needs. 

The energy release rate for the perturbed crack front is 
\beq{33}
G = PK = FK^2  = F^{-1} P^2.
\eeq
Here,  $P$ and $K$ are defined 
\beq{309}
P \equiv \eps^{-1/2}\lim \limits_{ X \uparrow 0} \left[ |X |^{-1/2} U_y (X,0, z, t)\right], 
\qquad
K \equiv \eps^{1/2}\lim \limits_{ X \downarrow 0} \left[ |X |^{1/2} \Sigma_{yy} (X,0, z, t)\right], 
\eeq
where $ U_y$ and $\Sigma_{yy}$ are the leading order term in the inner expansion of displacement, \rf{55b},  
and   stress, and \citep{Freund}
\beq{222}
F(V) = \frac{2\unota_L  (V) V^2}{v_T^2\mu D(V)}, 
\eeq
with $V=v+\eps f_t$. 
   Based on the lessons learned  from the  scalar problem, we  use the final expression in \rf{33} to find the perturbed 
energy release rate as
\beq{681}
\frac{\Delta G}{G_0}  =  2\frac{\Delta P}{P_0}  -\frac{F'(v)}{F(v)}f_t. 
\eeq

Following the procedure for the scalar problem, we can obtain  $\Delta P$  by matching the crack opening displacement. Thus, 
\beq{043}
U_y(-|X|,+0,z,t)= \frac{P_0}{\sqrt{2\pi}} \big[\eps^{1/2} (1+\eps p_1) |X|^{1/2} -
\eps^{3/2} \frac{m}{3} |X|^{3/2} \big]  + \O(\eps^{5/2}),
\eeq
and 
\beq{044}
U_y^{ \{ \frac32 , 1\} } (-|X|,+0,z,t)=  \frac{P_0}{\sqrt{2\pi}} \bigg[\big( |x|^{1/2}
-\frac{m}{3} |x|^{3/2} \big)
+\eps \,  \big[   \tfrac{1}{2} f\, |x|^{-1/2}  + \big(
p_1 -\frac{m}{2} f\big) |x|^{1/2}  \big]\bigg]  ,
\eeq
where now
\beq{045}
p_1 = B^{(3/2)} + \big[ \frac{3}{2\alpha_L\alpha_T}+\frac{(1+\alpha_T^2)v_T^2}{4\alpha_L^2v_L^2}\big]\frac{f_t}{v}
 = \eps^{-1} \frac{\Delta P}{P_0}. 
\eeq
The outer expansion for the crack opening is 
\beq{046}
u_y^{ \{1, \frac32\} } =  \frac{P_0}{\sqrt{2\pi}}  \big[ |x|^{1/2} -\frac{m}{3} |x|^{3/2} 
+\eps \, \Re e^{i(kz-\omega t)} 
\frac{\omega b_0}{2\pi} \int\limits_{-
\infty}^\infty \frac{\dd \xi\, e^{ i\omega \xi |x| }}{ (\xi -
i0)^{1/2} }  
\, \big( 1 + \frac{a}{\xi } +\O(\xi^{-2}) \big)
\big]  ,
\eeq
where the complex number $a$ is defined by the expansion of $\hat{u}_y^{(1)}$ in eqs.\ \rf{931} and \rf{304}.
Matching the the $|x|^{-1/2}$ singularity in the $\O(\eps)$ terms in \rf{044} and \rf{046} gives the identity \rf{404}, while the $|x|^{1/2}$ coefficient yields
\beq{604}
p_1 - \frac{m}2 f = 
\Im    \, i\omega a  f_0e^{i(k z- \omega t)}. 
\eeq
Hence, the transform for the perturbation in energy release rate, defined by 
$\rf{5.5}_3$, is
\beq{702}
\hat{g}(\omega, k) =  m +i \omega 2a + i \omega \frac{F'(v)}{F(v)} . 
\eeq
This is the fundamental dispersion relation, which will be discussed further in the next section.  
We note that a similar expression has been obtained by \cite[eq.\ (4.10)]{Willis01},  and by 
\citet[eq.\ (3.21)]{Woolfries02}\footnote{The parameter $m$ used here is the same as that of 
\citet{Woolfries02} but twice the value of the quantity $m$ in \citet{Willis01}.}.  However,  their expressions differ from   \rf{702} in the sign of the second and third terms.   This distinction is important for computing the dispersive form of the crack front wave (see section \rf{sec5}), but is immaterial in the  non-dispersive limit of $m/|k| \rightarrow 0$.  Finally, we note that the real parameter $B^{(3/2)}$ follows from eqs.\ \rf{045} and \rf{604}, and the complete inner expansion to $\O (\eps^{3/2})$ is then given by eq.\ \rf{7.12}.  

\begin{figure}[htbp]
				\begin{center}	
				\includegraphics[width=3.6in , height=2.8in 					]{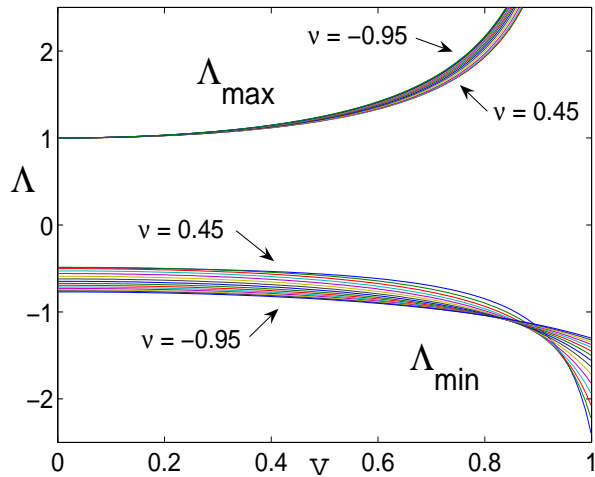} 
	\caption{The limits of the parameter $\Lambda$ of eq.\ \rf{266} as a function of the crack speed. The Poisson's ratio of the material ranges from $0.45$ to $-0.95$ in increments of $0.1$.  }
		\label{fig1} \end{center}  
	\end{figure}

\section{Numerical results and discussion}\label{sec5}

We discuss solutions to $\hat{g}(\omega, k) =0$ for  $0< v < 1$, where $v$ is the normalised crack speed relative to the Rayleigh wave speed.  Crack front waves with phase speed $c_p = \omega/k$ lie inside the Rayleigh  sonic cone if  
\beq{704}
c^2 = c_p^2+ v^2 < 1,
\eeq
where $c$ is the total speed of the disturbance in the fixed stationary frame. 
It may be shown \citep{Ramanathan97} that the transform $\hat{g}$ of \rf{702} is  real valued  for edge wave speeds inside the sonic cone.  Furthermore,  using the change of variables of
Ramanathan and Fisher  \cite{Ramanathan97}, also see \cite{Morrissey00}, we have 
\beq{703}
|k|^{-1}\hat{g}(\omega, k) =  \Lambda  
- 2\frac{\sqrt{1-c^2}}{1-v^2} +  \frac{\sqrt{1-c^2/v_L^2}}{1-v^2/v_L^2}
+    \int\limits_{v_T^2}^{v_L^2}\frac{\dd p}{\pi}
\, \frac{  [ 2v^2 p - c^2(v^2 + p)]\beta(p)}{ (p-v^2)^2 \sqrt{p(p-c^2)} },
\eeq
where 
\beq{7031}
 \beta(p) =
\tan^{-1} \bigr(
\frac{\sqrt{1-p/v_L^2}\sqrt{p/v_T^2 -1}}{\big(1-p/(2v_T^2)\big)^2}\bigr) , 
\eeq
and 
\beq{266}
\Lambda = |k|^{-1}m = \frac{\lambda m}{2\pi} . 
\eeq
Here $\lambda$ is the wavelength along the crack, and  $\Lambda$ is therefore the non-dimensional ratio of the wavelength to the length scale of the dynamic loading.  At a given crack speed, crack front waves   exist for a range of this parameter: 
$\Lambda_{\rm min}  < \Lambda < \Lambda_{\rm max} $, where the lower and upper limits correspond to $c=1$ and $c=v$, respectively.  Thus, 
\begin{subequations}\label{673}
\begin{align} 
\Lambda_{\rm min}(v)=&  - \frac{\sqrt{1-1/v_L^2}}{1-v^2/v_L^2}
-   \int\limits_{v_T^2}^{v_L^2}
  \frac{\dd p}{\pi}  \frac{( 2v^2 p - v^2 - p)\beta(p)}{ (p-v^2)^2 \sqrt{p(p-1)} } ,
\\
\Lambda_{\rm max}(v)=& 
\frac{2}{\sqrt{1-v^2}}  - \frac{1}{\sqrt{1-v^2/v_L^2}} 
-  \int\limits_{v_T^2}^{v_L^2}
 \frac{\dd p}{\pi}\frac{   v^2  \beta(p) }{ (p-v^2)^{3/2} \sqrt{p} } .
\end{align}
\end{subequations}
Figure \ref{fig1} show $\Lambda_{\rm min} $ and $ \Lambda_{\rm max}$ as a function of the crack speed for the range of elastic materials, characterised by the Poisson's ratio $\nu = (\frac12 v_L^2- v_T^2)/(v_L^2- v_T^2)$, which has the permissible range 
$-1<\nu < 0.5$.  We note that $\Lambda_{\rm min} $ is always negative and decreases monotonically to a finite value as $v\rightarrow 1$. 
Note also that $\Lambda_{\rm max} $ increases monotonically from unity at $v=0$ and goes as $\Lambda_{\rm max} \approx 2/\sqrt{1-v^2}$ as $v\rightarrow 1$.  This behaviour is to be compared with the scalar problem (see eq.\ \rf{5.61}) for which 
$\Lambda_{\rm min} =0$ and $\Lambda_{\rm max} =1/\sqrt{1-v^2}$.
The dependence upon Poisson's ratio is not strong, and therefore for the remainder we  consider the single case $\nu=1/3$.

\begin{figure}[htbp]
				\begin{center}	
				\includegraphics[width=3.6in , height=2.8in 					]{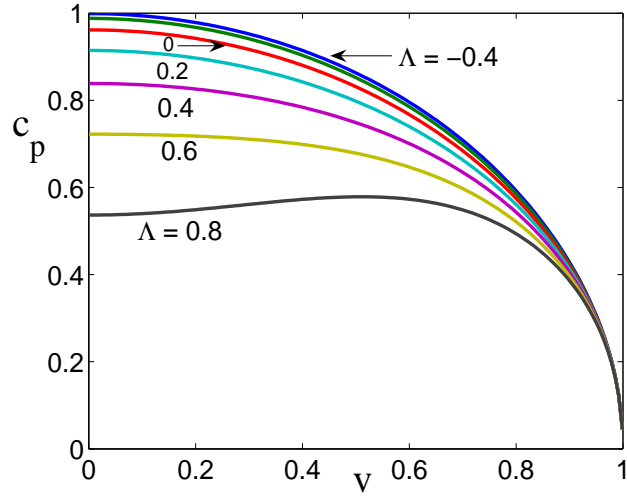} 
	\caption{The crack front phase speed $c_p$ for $\nu=1/3$ vs.\ the crack propagation speed for seven values of  the non-dimensional wavelength:  $\Lambda = -0.4$, $-0.2$,  $0$,  $0.2$, $0.4$, $0.6$ and $0.8$.  }
		\label{fig2} \end{center}  
	\end{figure}
\begin{figure}[htbp]
				\begin{center}	
				\includegraphics[width=3.6in , height=2.8in 					]{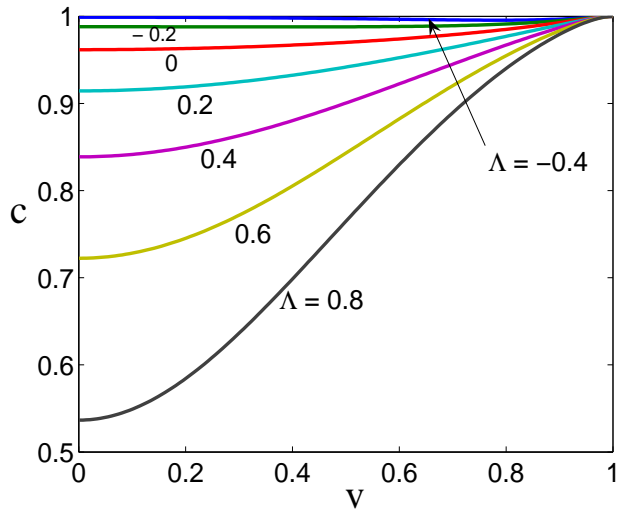} 
	\caption{The total speed of the crack front wave $c = \sqrt{v^2+c_p^2}$ vs.\ the crack propagation speed for seven values of  wavelength:  $\Lambda = -0.4$, $-0.2$,  $0$,  $0.2$, $0.4$, $0.6$ and $0.8$.  }
		\label{fig3} \end{center}  
	\end{figure}
\begin{figure}[htbp]
				\begin{center}	
				\includegraphics[width=3.6in , height=2.8in 					]{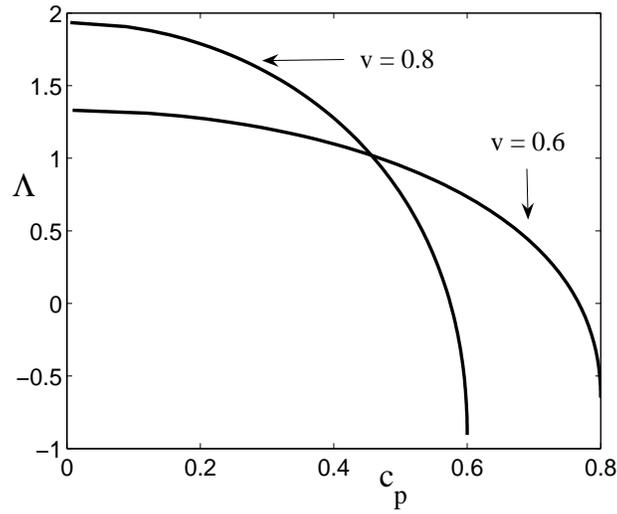} 
	\caption{The dependence of the crack front phase speed $c_p$ on the non-dimensional wavelength $\Lambda$ at two values of the crack speed, $v=0.6$ and $v=0.8$.  The upper limits of $c_p$ in each case are $\sqrt{1-v^2}$.  }
		\label{fig4} \end{center}  
	\end{figure}

The phase speed is plotted in Figure \ref{fig2} as a function of the crack speed for a range of the  non-dimensional wavelength $\Lambda$.  This extends the results of  Morrissey and Rice \cite{Morrissey00} who presented the $\Lambda=0$ curve for the total speed $c$, and  of Willis and Movchan \cite{Willis01} who computed  the $\Lambda =0$ curve for the phase speed. 
Figure \ref{fig3} shows the total velocity $c$ as a function of $v$ for the same set of $\Lambda$.  The total speed is always close to unity for negative values of $\Lambda$.  Thus, $0.995 < c < 1$,  $0.987 < c < 1$ and $0.961 < c < 1$ for $\Lambda = -0.4, -0.2$ and $0$, respectively.  Figure  \ref{fig4} plots the phase speed as a function of $\Lambda$ for two values of the crack speed.   The range of $\Lambda$ is consistent with Figure  \ref{fig1}, and note that the lower and upper limiting values of the phase speed correspond to $c=v$ and $c=1$, respectively.   
\begin{figure}[htbp]
				\begin{center}	
				\includegraphics[width=3.6in , height=2.8in 					]{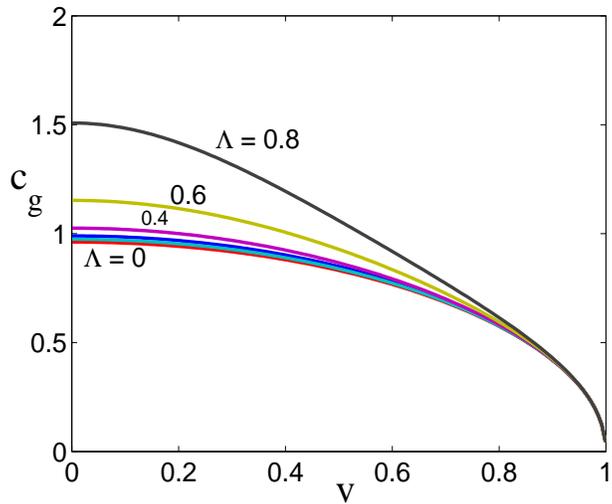} 
	\caption{The group speed $c_g$ of the crack front wave vs.\ $v$ for the same set of values for $\Lambda$ as in Figure \ref{fig3}.  }
		\label{fig5} \end{center}  
	\end{figure}

By analogy with the scalar problem, the group speed of the crack front wave travelling along the edge is defined $c_g=\dd \omega /\dd k$.  It may be shown to be 
\beq{73}
c_g =  c_p +  \frac{\Lambda}{c_p}\bigg( 
\frac{2}{(1-v^2)\sqrt{1-c^2}} - \frac{v_L}{(v_L^2-v^2)\sqrt{v_L^2-c^2}}
+  \int\limits_{v_T^2}^{v_L^2}\frac{\dd p}{\pi}
\, \frac{ ( c^2v^2 + c^2p - 2p^2)\beta(p)}{ (p-v^2)^2 (p-c^2)^{3/2}\sqrt{p} }\bigg)^{-1} ,
\eeq
and is plotted in Figure \ref{fig5} for a range of values of the wavelength-loading parameter $\Lambda$.  Note that values of $c_g$ in excess of unity occur, which is highly anomalous.  However, in contrast with the scalar case, the 
group speed does not always satisfy the inequality \rf{5.101}, although it holds more often than not .

\section{Conclusions}

We have shown how to solve the scalar model of a perturbed  travelling  crack edge by 
using the method of matched asymptotic expansions.  The  general procedure developed for 
the model elastic problem was then applied to the elastodynamic mode I crack.  The MAE 
solution employs  outer and inner expansions which are matched via the small parameter 
$\eps = L/l$ defined by the disparate length scales: the size $L$ of the edge  perturbation  
and the length scale $l$ associated with the loading.  The key to the solution is a  
homogeneous 
outer eigensolution which displays unphysical singularities at the unperturbed crack tip.  The reason for this is  that the crack is shifted, although  the unphysical eigensolution is not directly applicable at the crack edge.  No unphysical singularities are present in the inner solution, or in the complete matched solution.   We have used the MAE scheme to derive the dispersion relation for the crack front wave speed, and have shown its dependence on a range of parameters, including crack speed, wavelength, and elastic properties.  Once the MAE solution is found to second order in both the inner and outer fields, the dispersion relation  is obtained by requiring that the energy release rate, which depends on the inner solution,  is unaltered. 

The intent of this paper has been to show by example that MAE provides a natural methodology for solving problems of this nature.  Future work will develop the procedure to situations that are not easily solved by other means, for instance the perturbation of a dynamic crack front with a cohesive zone at the travelling crack tip.  In this case the size of the cohesive zone defines the inner length scale.

\appendix
\renewcommand{\theequation}{A.\arabic{equation}}
  \setcounter{equation}{0}  
  
\section{Solution of the $\O(\eps)$ outer problem }

We consider the symmetric  boundary 
value problem  defined by the  boundary conditions on $y=0$:
\begin{subequations}\label{12432}
\begin{align}
\sigma_{yy} = 0, &  \quad x < 0, \label{1a}
\\
u_y = 0, & \quad x >0,
\label{1b}
\\
\sigma_{xy} = \sigma_{zy} = 0, & \quad -\infty < x< \infty  . 
\label{1c}
\end{align}
\end{subequations}
This defines a mixed boundary value problems on 
the split regions $x <0$ and $x >0$.   The solution is assumed in the form \rf{1.05}, and 
 henceforth we will omit the 
term $e^{i\omega (\kappa z -t)}$ and concentrate on the transformed 
quantities $ \hat{\bf u}^{(1)}(\xi, y)$ and $ \hat{\bd{\sigma}}^{(1)}(\xi, 
y)$.

The symmetry of the problem implies that we need only consider the half space $y 
\ge 0$. The transform of the potentials   in this half-space may be represented by
\beq{2.22}
 \hat{\bd \phi}^{(1)} = (i\omega)^{-1}\begin{pmatrix}
  Ae^{-\omega \gamma_Ly}\\ Be^{-\omega \gamma_Ty} \\ (i\omega)^{-1}Ce^{-\omega \gamma_Ty}
\end{pmatrix}  ,
\eeq
and hence the displacement is  
\beq{a2.1}
 \hat{\bf u}^{(1)} (\xi, y)= A (-\xi, i\gamma_L, \kappa)^T 
 e^{-\omega \gamma_Ly}
+ \big[ B (i\gamma_T, \xi , 0)^T + C (-\xi \kappa, i\gamma_T \kappa , 
 \gamma_T^2-\xi^2 )^T\big] e^{-\omega \gamma_T y}, 
\eeq
where 
\beq{a2.2}
\gamma_L = \big( \xi^2 + \kappa^2 - \frac{1}{v_L^2}(1-
v\xi)^2\big)^{1/2}, \qquad
\gamma_T = \big( \xi^2 + \kappa^2 - \frac{1}{v_T^2}(1-
v\xi)^2\big)^{1/2}\, . 
\eeq
The square root functions are taken so that $\Re\gamma_L \ge 0$, 
 $\Re\gamma_T \ge 0$, thus ensuring decay in the half-space $y>0$.  

The associated stresses follow from eqs.\ \rf{1.07}  and 
\rf{a2.2}.  The transformed traction, when 
evaluated on the plane $y=0$, becomes
\beq{2.5}
(i\omega \mu)^{-1} \hat{\bd{\sigma}}^{(1)}(\xi, 0)= 
\begin{bmatrix}
-2i\gamma_L\xi  & - \xi^2 - \gamma_T^2  & -2i\gamma_T\xi \kappa \\  && 
\\
-(\xi^2 + \kappa^2 + \gamma_T^2) & 2i\gamma_T\xi & -2\gamma_T^2 \kappa \\  && 
\\
2i\gamma_L \kappa & \xi \kappa & i\gamma_T(  \kappa^2 + \gamma_T^2-\xi^2)
\end{bmatrix}
\begin{pmatrix}
A\\ \\ B\\ \\ C
\end{pmatrix}
. 
\eeq
It is also useful to list the analogous relation for the on-plane 
displacement transform, from \rf{a2.1} with $y=0$, 
\beq{2.6}
 \hat{\bf u}^{(1)}(\xi, 0)= 
\begin{bmatrix}
-\xi  &i\gamma_T  & -\xi \kappa \\  && \\
i\gamma_L  & \xi  & i\gamma_T\kappa \\  && \\
\kappa & 0 &  \gamma_T^2 - \xi^2
\end{bmatrix}
\begin{pmatrix}
A\\ \\ B\\ \\ C
\end{pmatrix}
. 
\eeq
We are now ready to consider the boundary conditions. 

The condition that the stresses $\sigma_{xy} $ and $\sigma_{zy} $ 
vanish on the plane $y=0$ implies that 
$\hat{\sigma}_{xy}^{(1)}(\xi, 0)=0 $ and $\hat{\sigma}_{zy}^{(1)}(\xi, 0)=0 $.  
The first and third rows of the vector relation \rf{2.5} then 
allows us to eliminate any two of $(A,B,C)$ in favour of the 
remaining quantity.  Substituting the result into the equations 
for $\hat{\sigma}_{yy}^{(1)}(\xi, 0)$ and $\hat{u}_y^{(1)} (\xi, 0)$, we 
deduce that 
\beq{a5.1}
\hat{\sigma}_{yy}^{(1)}(\xi, 0) = \frac{ \omega \mu v_T^2 R}{\gamma_L 
(1-v\xi)^2} \, 
 \hat{u}_y^{(1)} (\xi, 0)\, , 
\eeq
where $R$ is the (modified) Rayleigh function
\beq{a5.2}
R(\xi) = (\xi^2 + \kappa^2 + \gamma_T^2)^2 - 4 (\xi^2 + \kappa^2)\gamma_L 
\gamma_T\, . 
\eeq

The boundary conditions \rf{1a} and \rf{1b} imply that $\hat{\sigma}_{yy}^{(1)}(\xi, 0)$ is a $(+)$ function, 
i.e.\ analytic in the upper half of the complex $\xi$-plane, and $\hat{u}_y^{(1)} (\xi, 0)$ is a $(-)$ function. Define
\beq{104}
Q(\xi) = Q^+(\xi)Q^-(\xi) \equiv \frac{R(\xi)}{  (\eta_+ +\xi)(\eta_- - \xi)(\frac1{v} - \xi)^2 D}, 
\eeq
where $D$ is defined in \rf{6.4} and $\xi=-\eta_+$ and $\xi=\eta_-$ are the zeros of the modified Rayleigh function $R(\xi)$, i.e.\ the roots of 
$\xi^2 + \kappa^2 -(1-v\xi)^2 = 0$, 
\beq{121}
 \eta_\pm = \frac1{1-v^2}\big( \sqrt{1- \kappa^2 (1-v^2)} \pm v\big). 
\eeq
Thus, 
\beq{105}
Q(\xi) \rightarrow 1, \quad |\xi| \rightarrow \infty, 
\eeq
and the functions $Q^\pm (\xi)$ are defined unambiguously by  requiring that they both have this property.
Using the above mentioned analytic properties of the transforms we may rewrite \rf{a5.1} as
\beq{32}
\frac{\gamma_L^+(\xi) \hat{\sigma}_{yy}^{(1)}(\xi, 0) } {Q^+ (\xi)(\eta_+ +\xi) }
 = \frac{ \omega \mu v_T^2 D}{v^2 \gamma_L^-(\xi) } \, 
Q^- (\xi)(\eta_- -\xi) \hat{u}_y^{(1)} (\xi, 0) \equiv  E(\xi) , 
\eeq
where, 
by the usual analytic continuation arguments, $E$ is an entire function of the complex variable $\xi$ and 
\beq{4.71}
 \gamma_L^{\pm}(\xi) = \unota_L\, \big( \xi \pm \lambda_{L\pm}\big)^{1/2}, \qquad
\lambda_{L\pm} = \frac{1}{v_L\unota_L^2} \big( 1-\kappa^2 v_L^2 \unota_L^2\big)^{1/2} \pm \frac{v}{v_L^2\unota_L^2}
\, .
\eeq
Assume that the stress and displacement behave as 
\beq{74}
{u}_y^{(1)} (x, 0) = \O( (-x)^{-1/2}), 
\qquad
\sigma_{yy}^{(1)}(x, 0) = \O( (-x)^{-3/2}), 
\eeq
near the edge.  This implies that the transforms behave as $\xi^{-1/2}$ and $\xi^{1/2}$ as $\xi $ tends to infinity, and therefore the entire function $E$ is in fact a constant.  We assume the constant is non-zero otherwise ${\bd \phi}^{(1)} =0$ and the matching cannot be accomplished.  

We note that the only non-trivial solution to \rf{a5.1} has an unphysical 
singularity near the edge of the crack.  This removes  the 
possibility of infinitesimal localised disturbances, analogous to 
Rayleigh waves on a free surface. Achenbach and Gautesen \cite{Achenbach77} considered the possible 
existence of travelling waves on the edge of a stationary crack, 
and showed that neither symmetric nor anti-symmetric solutions can 
exist.  Equation \rf{a5.1}  
generalises their   result to steady 
propagating cracks, and shows the non-existence of symmetric 
localized modes.  A similar analysis can be done to show that 
anti-symmetric modes cannot exist on the propagating crack.  

In general, the near tip behaviour follows from the expansion of the potentials as $|\xi|\rightarrow \infty$.  
Using 
\beq{303}
Q^-(\xi) = \exp\bigg( \frac{-1}{2\pi i} \int\limits_{-\infty}^\infty \dd \zeta \, \frac{ \ln Q(\zeta)}{\zeta - \xi}\bigg), 
\eeq
we have 
\beq{931}
 \hat{u}_y^{(1)}(\xi, 0)  = \frac{b_0}{\xi^{1/2}}\big[ 1 + \frac{a}{\xi} + \O(\xi^{-2})\big], 
\eeq
where, from \rf{32} and \rf{4.71}, 
\beq{304}
a =  \eta_- - \frac{1}{2} \lambda_{L-} - \frac{1}{2\pi i} \int\limits_{-\infty}^\infty \dd \zeta   \ln Q(\zeta) . 
\eeq
This is used in section \rf{sec4} to obtain the perturbed energy release rate. The detailed form of the inner expansion and the near tip field can be obtained by noting that the  explicit form of the potentials are  
\beq{a9}
\begin{pmatrix}
A\\ \\ B\\ \\ C
\end{pmatrix}
= \frac{ 
 \hat{u}_y^{(1)} (\xi, 0)}{ i\,\gamma_L  (    \gamma_T^2 - \xi^2 )(  
\xi^2 +\kappa^2-  \gamma_T^2) }\, 
\begin{pmatrix}
(\xi^2 + \gamma_T^2)(  \xi^2 -\kappa^2-  \gamma_T^2) +2\xi^2 \kappa^2\\   \\
-2i\xi \gamma_L  (  \xi^2 +\kappa^2-  \gamma_T^2) \\   \\
2\kappa \gamma_L\gamma_T 
\end{pmatrix}
\eeq
which imply 
\begin{subequations} \label{723}
\begin{align}\label{081}
A&= (1+\unota_T^2)|\xi|^{-1}\big( 
1 + \frac{1}{v\xi} \big[ \frac4{1+\unota_T^2} - \frac{(1-\unota_L^2)}{\unota_L^2}\big]
+\ldots \big) \, \hat{u}_y^{(1)}(\xi, 0) , 
\\
B&= -2i \unota_L \xi^{-1}\big( 
1 + \frac{2}{v\xi} 
+\ldots \big) \, \hat{u}_y^{(1)}(\xi, 0) , 
\\
C&= \frac{2  \unota_L\unota_T \kappa }{1-\unota_T^2}
|\xi|^{-3} \big( 
1 + \frac{1}{v\xi} \big[ \frac1{\unota_T^2}  + 3\big] 
+\ldots \big) \, \hat{u}_y^{(1)}(\xi, 0) . 
\end{align}
\end{subequations}
Equations \rf{931} and \rf{723} can then be used to derive the near-tip fields. 



\end{document}